\documentclass{article}

\usepackage{microtype}
\usepackage{graphicx}
\usepackage{subfigure}
\usepackage{booktabs} % for professional tables

\usepackage{hyperref}

\usepackage[accepted]{icml2025}

\usepackage{amsmath}
\usepackage{amssymb}
\usepackage{mathtools}
\usepackage{amsthm}

\usepackage[capitalize,noabbrev]{cleveref}

\usepackage{listings}
\usepackage{tcolorbox}
\usepackage{pgfplots}
\pgfplotsset{compat=1.18}
\usetikzlibrary{chains,positioning,arrows.meta,shapes.geometric,calc}
\usepackage{enumitem}
\usepackage{multirow}

\theoremstyle{plain}

\theoremstyle{definition}

\theoremstyle{remark}

\newtcolorbox{insightbox}[1][]{%
  colback=blue!5!white,
  colframe=blue!50!white,
  colbacktitle=blue!50!white,
  leftrule=4pt,
  rightrule=0.5pt,
  toprule=0.5pt,
  bottomrule=0.5pt,
  fonttitle=\bfseries,
  #1
}

\newtcolorbox{insightwarm}[1][]{%
  colback=orange!5!white,
  colframe=orange!60!white,
  colbacktitle=orange!60!white,
  leftrule=4pt,
  rightrule=0.5pt,
  toprule=0.5pt,
  bottomrule=0.5pt,
  fonttitle=\bfseries,
  #1
}

\newtcolorbox{insightgreen}[1][]{%
  colback=green!5!white,
  colframe=green!50!black,
  colbacktitle=green!50!black,
  leftrule=4pt,
  rightrule=0.5pt,
  toprule=0.5pt,
  bottomrule=0.5pt,
  fonttitle=\bfseries,
  #1
}

\newtcolorbox{gatebox}[1][]{%
  colback=yellow!5!white,
  colframe=orange!70!yellow,
  colbacktitle=orange!70!yellow,
  leftrule=4pt,
  rightrule=0.5pt,
  toprule=0.5pt,
  bottomrule=0.5pt,
  fonttitle=\bfseries,
  #1
}

\icmltitlerunning{KAIJU: An Executive Kernel for Intent-Gated Execution of LLM Agents}

\begin{document}

\twocolumn[
\icmltitle{KAIJU: An Executive Kernel for Intent-Gated Execution of LLM Agents}

\icmlsetsymbol{equal}{*}

\begin{icmlauthorlist}
\icmlauthor{Cormac Guerin}{comp}
\icmlauthor{Frank Guerin}{sch}
\end{icmlauthorlist}

\icmlaffiliation{comp}{Compdeep}
\icmlaffiliation{sch}{School of Computer Science and Electronic Engineering, University of Surrey, UK}

\icmlcorrespondingauthor{Cormac Guerin}{cormac@compdeep.com}
\icmlcorrespondingauthor{Frank Guerin}{f.guerin@surrey.ac.uk}

\icmlkeywords{Agentic AI}

\vskip 0.3in
]

\printAffiliationsAndNotice{}

\begin{abstract}
Tool-calling autonomous agents based on large language models 
%in ReAct configurations 
using ReAct
exhibit three %compounding 
limitations: serial latency, quadratic context growth, and vulnerability to prompt injection and hallucination. %; collectively producing outcomes that are slow, costly, or potentially harmful, particularly acute in domains such as cybersecurity and defence.
Recent work moves towards separating planning from execution but in each case the model remains coupled to the execution mechanics. We introduce a system-level abstraction for LLM agents which decouples the execution of agent workflows from the LLM reasoning layer. We define two first-class abstractions: (1) Intent-Gated Execution (IGX), a security paradigm that enforces intent at execution, and (2) an Executive Kernel that manages scheduling, tool dispatch, dependency resolution, failures and security.
 %Consisting of an independent directed acyclic graph, the LLM plans upfront, optimistically scheduling tools in parallel via a dependency-aware execution graph with parameter injection; lightweight reflection nodes evaluate intermediate results and guide the graph toward resolution.
 In KAIJU, the LLM plans upfront, optimistically scheduling tools in parallel with dependency-aware parameter injection.
 Tools are authorised via IGX based on four independent variables: scope, intent, impact, and clearance (external approval). 
 KAIJU supports three adaptive execution modes (Reflect, nReflect, and Orchestrator), providing progressively finer-grained execution control apt for complex investigation and deep analysis or research. 
Empirical evaluation against a ReAct baseline %operating over the same tools and safety gates 
shows that KAIJU has a latency penalty on simple queries due to planning overhead, convergence at moderate complexity, and a structural advantage on computational queries requiring parallel data gathering. Beyond latency, the separation enforces behavioural guarantees that ReAct cannot match through prompting alone. %: the execution layer persists through tool failures via automatic replanning, produces bounded-context LLM calls that avoid cumulative degradation, and delegates reflection to a cheaper executor model while ReAct must use the reasoning model for every turn.
\\
Code  available at \href{https://github.com/compdeep/kaiju}{github.com/compdeep/kaiju}
\end{abstract}

%% }{======================================================================
%% 1. INTRODUCTION
%% ============================================================================

\section{Introduction}
\label{sec:introduction}

LLM-based agents that invoke external tools (APIs, shell commands, databases, web services) have become a dominant paradigm for grounding language models in real-world action. The foundational execution model is ReAct \citep{yao2023react}: the model reasons about which tool to call, executes it, observes the result, and repeats. Modern implementations in production systems such as OpenAI's Assistants API, Anthropic's Claude Code, and OpenClaw extend this with native parallel function calling; the model can return multiple tool calls per turn, significantly reducing the number of reasoning round-trips. On straightforward queries, this approach is fast and effective.

Three problems emerge as task complexity grows. First, each reasoning turn still carries the full conversation history. Turn $i$ transmits $c + ik$ tokens, where $c$ is the system context and $k$ is the average tool result size. Total cost across $n$ turns is $O(n^2 k)$; for typical values of $k \approx 1$K tokens, a 7-tool task transmits approximately 63K tokens; an 18-tool task approximately 250K. On multi-step computational queries, this cumulative context can exhaust the effective processing window, producing empty or degraded outputs. Second, the model retains unilateral authority over tool use at every turn. When a tool call fails or returns partial results, the model can abandon the task, defer to parametric knowledge, or ask the user for guidance; behaviour that is rational per-turn but undermines reliability across the full query. Prompt instructions to persist (``never give up, always retry'') partially mitigate this but cannot guarantee it; the model may comply or not on each turn. Third, tool safety is enforced via prompt instructions (``do not call destructive tools''), which the model can ignore through hallucination, prompt injection, or context overflow. There is no structural failsafe. \citet{nasr2025attacker} demonstrate that LLM-level defenses whether training-based, prompt-based, or filter-based are systematically defeated by adaptive adversaries who modify their attack strategy against the defense's observable behaviour.

Prior systems address these problems partially. LLM Compiler \citep{kim2024llmcompiler} introduced parallel DAG execution but evaluates results only after the entire graph completes, with no mid-execution adaptation and no safety gating. LangGraph \cite{langgraph2024,biju2025langgraph} models workflows as state machines but keeps the LLM in the execution loop at every routing decision, with safety limited to human-in-the-loop breakpoints. Multi-agent frameworks (\citet{crewai2024}, AgentArch \cite{agentarch2025}) add coordination overhead that degrades both performance and reliability; \citet{agentarch2025} found that even the best configurations achieved only 35.3\% success on complex enterprise tasks with 6.3\% reproducibility. In each case, the model remains coupled to execution semantics.

We introduce KAIJU, with a strict execution abstraction consisting of two layers: a reasoning layer handling user interaction, and an execution layer responsible for dependency resolution, tool dispatch, failure recovery, safety enforcement, and result synthesis. In the execution layer the LLM is a stateless resource invoked at discrete points (to plan, to reflect, and to aggregate) with no visibility into execution mechanics. The execution layer invokes the reasoning model once to produce a dependency graph of tool calls, then schedules, gates, and dispatches tools independently, constituting the first graph node. At structured leaf nodes, a lightweight reflector evaluates the evidence gathered so far against the original query and decides whether to continue, conclude, or replan with targeted follow-up nodes. An execution gate (IGX) enforces authorisation via four independent variables (scope, intent, impact, and clearance), each governed by a separate authority outside the model.

The LLM does not observe gate decisions within the execution layer, removing the possibility of adaptive probing of the safety policy. In conversational loops (ReAct, LangGraph), blocked tool calls feed back into the model's context, enabling adversarial probing; our architecture removes the feedback channel. Critically, data are only input to the execution layer from the reasoning layer at call time and fed back only at return time, creating a closed execution loop, except for preemption as described in Section~\ref{preempt}. The execution layer persists through tool failures via automatic replanning rather than deferring to the user. It does not ask for permission or silently substitute parametric knowledge; when a tool fails, the micro-planner retries with alternative approaches before the reflector evaluates whether sufficient evidence has been gathered. The graph is processed optimistically (execute first, adapt on failure via reflection checkpoints) across three adaptive modes. 

A structured graph-based LLM workflow enables composable, verifiable execution units, providing a foundation for enforcing guarantees such as security and content policy and service level agreements. Nodes can be added or removed to improve or alter the graph mid-flight enabling custom and controlled outcomes. In our system, three execution modes provide progressively finer-grained mid-execution adaptation: structural phase boundaries (Reflect), periodic batch checkpoints (nReflect), and per-node observers (Orchestrator).

\textbf{Contributions of IGX:}
\vspace{-1ex}
\begin{enumerate}[leftmargin=*]
  \item \textbf{Separation + bounded context $\rightarrow$ reduced token scaling.} The planner and per-node observers operate on bounded context. Reflections see cumulative evidence within the execution layer but not the full conversational history. Token complexity reduces from $O(n^2 k)$ to $O(nkd)$ in Reflect mode or $O(nk)$ in Orchestrator mode, where $d$ is dependency depth.

  \item \textbf{Separation + parallelism $\rightarrow$ schedulable.} The execution graph fires tools on dependency resolution, not LLM decision. Three adaptive modes (Reflect, nReflect, Orchestrator) reduce latency to $O(d)$ where $d$ is dependency depth, compared to $O(n)$ for sequential dispatch or $O(t)$ for parallel function calling (where $t$ is reasoning turns, $t \ll n$ but structurally unbounded).

  \item \textbf{Separation + enforcement $\rightarrow$ structurally enforced safety.} A four-variable gate (scope, intent, impact, clearance) enforces authorisation deterministically in compiled code. Gate decisions do not reach the model; it does not distinguish a blocked tool from a failed one, preventing adaptive iteration against the policy.

  \item \textbf{Separation + dataflow $\rightarrow$ structural dependency injection.} A \texttt{param\_refs} mechanism expresses data flow between steps at plan time; concrete values resolve at execution time from upstream outputs. This removes the need for a sequential reasoning loop to pass data between tool calls.

  \item \textbf{Delegated resource clearance.} Resource-level authorisation delegates to external HTTP endpoints, enabling deployment across cybersecurity, robotics, enterprise, and healthcare environments without domain-specific logic in the agent.
\end{enumerate}

The remainder of this paper is organised as follows. Section~\ref{sec:related} surveys related work. Section~\ref{sec:example} introduces an example illustrating key properties. Section~\ref{sec:model} defines the system model and graph structure. Section~\ref{sec:modes} describes the three execution modes. Section~\ref{sec:gate} presents the execution gate. Section~\ref{sec:experiments} reports experimental results. Section~\ref{sec:conclusions} concludes with limitations and future directions.

%% ============================================================================
%% 2. RELATED WORK
%% ============================================================================

\section{Related Work}
\label{sec:related}

%\subsection{Sequential agent execution}

We give an overview of the principal related works in  Table~\ref{tab:related}.
ReAct \citep{yao2023react} established sequential think-act-observe with $O(n^2 k)$ token complexity. In its original formulation, latency is $O(n)$ with one tool per turn; modern implementations with parallel function calling (OpenAI, Anthropic) reduce this to $O(t)$ reasoning turns, though token complexity remains quadratic. Each turn appends tool results to a growing context, degrading both cost and attention quality on longer tasks. AutoGPT \cite{autogpt2023} and BabyAGI \cite{babyagi2023} extended this with persistent memory but retained the conversational execution model. All enforce safety through prompt instructions only.

%\subsection{Parallel and graph-based execution}

LLM Compiler \citep{kim2024llmcompiler} is the closest prior work on parallel execution. It generates a DAG and dispatches tool calls in parallel, reporting up to 3.7$\times$ speedup compared to sequential ReAct. In our testing against squential ReAct we observe the same gain however on complex queries 7x and on highly complex queries up to 18x increase. LLM Compiler  evaluates results only after the entire graph completes (no mid-execution adaptation), using whole-output substitution rather than field-level dependency injection, provides no safety gating is limited to human-in-the-loop breakpoints. By contrast our system extends this four degrees; dependency injection (vs.\ whole-output substitution), mid-execution adaptation via three modes (vs.\ post-completion Joiner only), structural safety gating (vs.\ none), and scoped failure recovery (vs.\ full replanning).
\begin{itemize}
\item LLM Compiler: all tools execute → Joiner evaluates → if insufficient, replan entire graph → execute again. Adaptation is synchronous and epoch-based. Nothing changes while tools are running.
\item KAIJU Executive Kernel: tools execute in waves → reflection can fire between waves while other branches may still be pending → micro-planner can graft new nodes onto a live graph → observers (Orchestrator mode) evaluate individual results as they arrive and inject follow-up nodes immediately. The graph is mutable during execution.
\end{itemize}

%\subsection{Multi-agent frameworks}

\citet{crewai2024} and AgentArch \cite{agentarch2025}
 organise agents by role. \citet{agentarch2025} found that multi-agent configurations ``consistently underperformed'' single-agent function calling, achieving only 35.3\% success on complex enterprise tasks with 6.3\% reproducibility. The system achieves specialisation through the tool interface rather than agent personas, avoiding the coordination overhead that degrades multi-agent reliability.

%\subsection{LLM-level safety defenses}

Most agent safety approaches operate within the LLM's own processing: safety training, input/output filtering, guardrail prompts, or fine-tuned refusal behaviour. \citet{nasr2025attacker} systematically evaluate 12 such defenses against adaptive adversaries and breach all with ${>}90$\% success rates, demonstrating that defenses observable to the model are fundamentally vulnerable to iteration. With KAIJU IGX operates outside the LLM's control and observation surface; the model produces a plan but never sees whether its tool calls were permitted, denied, or gated.

%\subsection{Structural tool control}

CaMeL \citep{debenedetti2025camel} addresses prompt injection by separating control flow from data flow, preventing untrusted inputs from influencing program execution. It achieves provable guarantees on 77\% of AgentDojo tasks. CaMeL focuses on data exfiltration prevention (what information flows where); our work focuses on tool execution control (which tools execute at what impact level). The two are complementary and composable.

Progent \citep{shi2025progent} provides programmable privilege control via argument-level condition matching, achieving 0\% attack success on AgentDojo. Policies are expressed as boolean conditions over tool arguments (regex, comparisons, membership). Key differences: Progent operates within the conversational loop (blocked calls feed back to the agent as messages, enabling probing of the policy boundary), requires manual or LLM-generated policy authoring per tool argument, and does not support dynamic per-invocation impact classification or external authorisation. Our gate uses four scalar variables rather than arbitrary boolean expressions, and is structurally isolated from the LLM's observation.

\begin{table*}[!ht]
\caption{Comparison of agent execution systems.}
\label{tab:related}
\vskip 0.15in
\begin{center}
\begin{small}
\begin{tabular}{@{}llll@{}}
\toprule
System & Execution & Safety mechanism & LLM sees rejection? \\
\midrule
ReAct & Sequential & Prompt instructions & Yes \\
LangGraph & State machine & Human-in-loop & Yes \\
LLM Compiler & DAG & None & N/A \\
CaMeL & Sequential & Data flow sep. & No (ctrl plane) \\
Progent & Sequential & Arg.\ cond.\ matching & Yes (feedback) \\
\textbf{IGX (ours)} & \textbf{DAG (3 modes)} & \textbf{4-variable\ intent gate} & \textbf{No (structural\ separation)} \\
\bottomrule
\end{tabular}
\end{small}
\end{center}
\vskip -0.1in
\end{table*}

%% ============================================================================
%% 3. RUNNING EXAMPLE
%% ============================================================================

\section{Running Example}
\label{sec:example}

We illustrate with the following query: \emph{``Check disk usage, list open ports, find git repos under /home, search for CVEs affecting our kernel version, and check environment variables for leaked secrets.''} This requires five independent data-gathering operations, a dependency chain (kernel version must be discovered before searching for CVEs), and cross-domain synthesis. Both queries were run through the same system instance with identical tools. 

\paragraph{ReAct with parallel function calling --- 9 turns, 14 tools, 64.5s.}

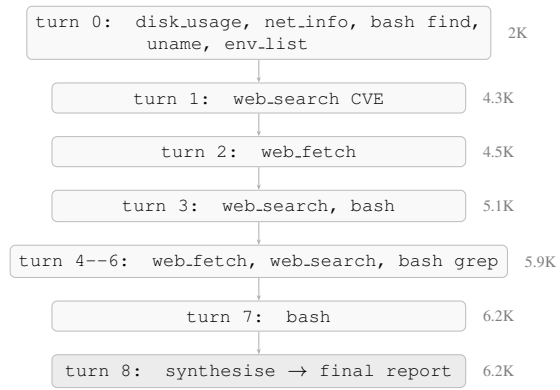
\begin{center}
\begin{tikzpicture}[
  node distance=0.3cm,
  turn/.style={rounded corners=2pt, draw=gray!50, fill=gray!5, font=\scriptsize\ttfamily, minimum width=5.5cm, minimum height=0.4cm, align=left},
  ctx/.style={font=\tiny\color{gray}, anchor=west},
  >={Stealth[length=2pt]}
]
\node[turn] (t0) {turn 0: disk\_usage, net\_info, bash find,\\        \phantom{turn 0: }   uname, env\_list};
\node[ctx, right=0.1cm of t0] {2K};
\node[turn, below=of t0] (t1) {turn 1: web\_search CVE};
\node[ctx, right=0.1cm of t1] {4.3K};
\node[turn, below=of t1] (t2) {turn 2: web\_fetch};
\node[ctx, right=0.1cm of t2] {4.5K};
\node[turn, below=of t2] (t3) {turn 3: web\_search, bash};
\node[ctx, right=0.1cm of t3] {5.1K};
\node[turn, below=of t3] (t4) {turn 4--6: web\_fetch, web\_search, bash grep};
\node[ctx, right=0.1cm of t4] {5.9K};
\node[turn, below=of t4] (t7) {turn 7: bash};
\node[ctx, right=0.1cm of t7] {6.2K};
\node[turn, below=of t7, fill=gray!15] (t8) {turn 8: synthesise $\rightarrow$ final report};
\node[ctx, right=0.1cm of t8] {6.2K};
\foreach \a/\b in {t0/t1,t1/t2,t2/t3,t3/t4,t4/t7,t7/t8}
  \draw[->, gray!60] (\a) -- (\b);
\end{tikzpicture}
\end{center}

Modern parallel function calling batches 5 tools on turn 0, reducing the original 12-turn sequential baseline to 9 turns. However, each subsequent turn still carries the full conversation history; by turn 8 the context exceeds 6K tokens. The model decides per-turn whether to continue researching or conclude; turns 1--7 show it pursuing CVE details across multiple sequential search-fetch cycles. Total: 9 LLM calls, 14 tool executions, 64.5 seconds.

\newpage

\paragraph{IGX (DAG, Reflect) --- 4 LLM calls, 10 nodes, 41.8s.}

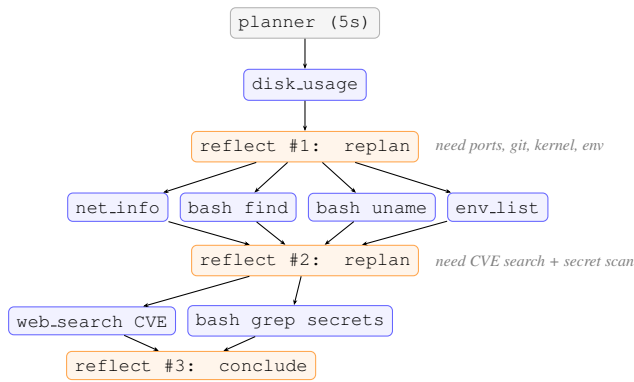
\begin{center}
\begin{tikzpicture}[
  node distance=0.25cm and 0.15cm,
  tool/.style={rounded corners=2pt, draw=blue!60, fill=blue!5, font=\scriptsize\ttfamily, minimum height=0.35cm, inner sep=3pt},
  refl/.style={rounded corners=2pt, draw=orange!70, fill=orange!8, font=\scriptsize\ttfamily, minimum height=0.35cm, inner sep=3pt},
  plan/.style={rounded corners=2pt, draw=gray!60, fill=gray!8, font=\scriptsize\ttfamily, minimum height=0.35cm, inner sep=3pt},
  note/.style={font=\tiny\color{gray}, anchor=west},
  >={Stealth[length=2pt]}
]
% Wave 0
\node[plan] (pln) {planner (5s)};
\node[tool, below=0.4cm of pln] (du) {disk\_usage};
\draw[->] (pln) -- (du);

% Reflect 1
\node[refl, below=0.4cm of du] (r1) {reflect \#1: replan};
\node[note, right=0.1cm of r1] {\textit{need ports, git, kernel, env}};
\draw[->] (du) -- (r1);

% Wave 1 — parallel
\node[tool, below left=0.4cm and 0.3cm of r1] (ni) {net\_info};
\node[tool, right=0.15cm of ni] (bf) {bash find};
\node[tool, right=0.15cm of bf] (bu) {bash uname};
\node[tool, right=0.15cm of bu] (el) {env\_list};
\draw[->] (r1) -- (ni); \draw[->] (r1) -- (bf); \draw[->] (r1) -- (bu); \draw[->] (r1) -- (el);

% Reflect 2
\node[refl, below=0.5cm of $(bf)!0.5!(bu)$] (r2) {reflect \#2: replan};
\node[note, right=0.1cm of r2] {\textit{need CVE search + secret scan}};
\draw[->] (ni) -- (r2); \draw[->] (bf) -- (r2); \draw[->] (bu) -- (r2); \draw[->] (el) -- (r2);

% Wave 2 — parallel
\node[tool, below left=0.4cm and 0.2cm of r2] (ws) {web\_search CVE};
\node[tool, right=0.15cm of ws] (bg) {bash grep secrets};
\draw[->] (r2) -- (ws); \draw[->] (r2) -- (bg);

% Reflect 3
\node[refl, below=0.4cm of $(ws)!0.5!(bg)$] (r3) {reflect \#3: conclude};
\draw[->] (ws) -- (r3); \draw[->] (bg) -- (r3);
\end{tikzpicture}
\end{center}

The planner under-planned on this query, producing only one node initially. The reflection loop compensated: three reflections identified gaps and replanned with targeted follow-ups across two additional waves. Despite this, the DAG completed in 41.8 seconds versus ReAct's 64.5 seconds because each wave executed tools in parallel and each reflection saw only the current wave's results (bounded context) rather than the full conversation history. The CVE search in wave 3 used the kernel version discovered in wave 2. Total: 4 LLM calls (planner + 3 reflections), 10 tool executions, 41.8 seconds. The final reflection produced the verdict directly, skipping the aggregator.

\begin{insightwarm}
\textbf{The running example illustrates three key properties.} (1) Parallel execution within dependency waves reduces wall clock time even when the planner under-plans; the reflection loop compensates by identifying gaps and launching targeted follow-up waves. (2) Bounded context per LLM call: each reflection evaluates only the current wave's evidence, not all prior tool results, avoiding the cumulative context growth visible in the ReAct trace (2K $\rightarrow$ 6.2K). (3) The execution layer persists through incomplete plans via structural replanning; the model cannot decide to stop early or defer to the user.
\end{insightwarm}

%% ============================================================================
%% 4. SYSTEM MODEL
%% ============================================================================

\section{System Model}
\label{sec:model}

The IGX system executes discrete operations as \textbf{nodes} within a directed acyclic graph. Every node carries a unique identifier, a type, a lifecycle state, upstream dependencies, a parameter map, and a result slot populated upon completion. Nodes are connected by directed edges through \texttt{depends\_on} references: a node will not fire until every dependency has reached a terminal state.

The planner LLM produces the initial graph from the task description. During execution, the graph is \textbf{live}; reflection checkpoints, observers, and failure-repair calls may graft additional nodes at any point. Every state transition is emitted as a streaming event, making the full execution trace observable in real time. This liveness is central to the system's adaptive capability: the graph at completion may differ substantially from the graph the planner originally produced.

\subsection{Node Types}

The system defines six node types:

\begin{itemize}[leftmargin=*]
  \item \textbf{Tool.} Executes a registered tool with declared parameters. Subject to the execution gate. The primary node type.
  \item \textbf{Reflection.} LLM checkpoint that evaluates completed results. Can continue, conclude early, or replan by grafting new nodes.
  \item \textbf{Observer.} Lightweight per-node LLM call (Orchestrator mode). Evaluates a single result; can inject nodes, cancel pending work, or trigger reflection.
  \item \textbf{Micro-Planner.} Scoped failure-repair call. Evaluates a failed node and produces replacement nodes: retry with different parameters, substitute an alternative tool, or skip.
  \item \textbf{Interjection.} Human-triggered checkpoint. The operator's message gates all pending nodes, giving the ability to redirect or halt an active execution.
  \item \textbf{Aggregator.} Optional final LLM call. Synthesises all gathered results into a structured output. Fires only after all other nodes complete and the final reflection concludes. Can be disabled (the reflection verdict is used directly), run on the executor model for speed, or on the reasoning model for quality. Disabled for all benchmarks in this paper.
\end{itemize}

\subsection{Node Lifecycle}

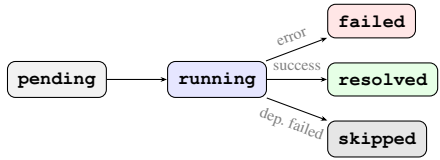
\begin{center}
\begin{tikzpicture}[
  state/.style={rounded corners=3pt, draw, font=\scriptsize\ttfamily\bfseries, minimum height=0.4cm, inner sep=4pt},
  note/.style={font=\tiny\color{gray}},
  >={Stealth[length=2.5pt]}
]
\node[state, fill=gray!10] (p) {pending};
\node[state, fill=blue!10, right=0.8cm of p] (r) {running};
\node[state, fill=green!10, right=0.8cm of r] (ok) {resolved};
\node[state, fill=red!10, above right=0.3cm and 0.8cm of r] (f) {failed};
\node[state, fill=gray!20, below right=0.3cm and 0.8cm of r] (s) {skipped};
\draw[->] (p) -- (r);
\draw[->] (r) -- (ok) node[note, midway, above] {success};
\draw[->] (r) -- (f) node[note, midway, above, sloped] {error};
\draw[->] (r) -- (s) node[note, midway, below, sloped] {dep.\ failed};
\end{tikzpicture}
\end{center}

A node remains \textbf{pending} until all dependencies reach a terminal state. It then transitions to \textbf{running}, at which point parameter injections are resolved and the operation executes. Terminal states are final; a node is not re-executed. If a node fails, the micro-planner may graft replacement nodes rather than retrying the original. This immutability simplifies reasoning about the graph: once a node reaches a terminal state, its result is stable and available to all downstream consumers.

\subsection{Edges: Ordering and Data Flow}

\textbf{\texttt{depends\_on}} establishes execution ordering. A node will not fire until all referenced nodes have completed. Nodes without dependencies execute immediately and in parallel. This creates a natural wave structure: all root nodes (those with no dependencies) fire concurrently in the first wave, and subsequent waves fire as their dependencies resolve.

\textbf{\texttt{param\_refs}} establishes data flow. Each entry maps a parameter name to a source node, a dot-path field selector into that node's JSON result, and an optional template string. At fire time, the scheduler extracts the value, applies the template if present, and injects the result into the parameter map. Any node referenced in \texttt{param\_refs} is automatically added to \texttt{depends\_on}, ensuring that data dependencies are always respected even if the planner omits the explicit ordering edge.

\begin{lstlisting}[basicstyle=\scriptsize\ttfamily]
Example: meeting booking with dependency injection

step 0  check_calendars
  params: {team: "engineering"}
step 1  create_event
  params: {title: "Sync"}
  param_refs: {start: {step:0, field:"slots.0.start"}}
  depends_on: [0]
step 2  send_invites
  param_refs: {event_id: {step:1, field:"event_id"},
               attendees: {step:0, field:"members"}}
  depends_on: [0, 1]
\end{lstlisting}

Each value materialises exactly when it becomes available. The planner declared the complete three-step workflow upfront without knowing any concrete values; it only specified \emph{where} each value would come from. The mechanism is structural (the scheduler enforces dependency ordering), fails fast (missing fields produce immediate diagnostic errors), and composable (a single step may reference multiple source nodes). This bridges the gap between upfront planning and runtime data availability without collapsing to sequential execution.

\subsection{Iterative Final Reflection}

A structural guarantee ensures that a reflection evaluates results after every batch of tool completions. When all planned nodes complete, if skill work has occurred since the last reflection, the system injects a new reflection node into the graph. If the reflection replans, the new nodes execute through the normal scheduler loop, and the process repeats. This continues until the reflection concludes, the budget is exhausted, or the wall clock expires. The aggregator fires only after the final reflection approves.

This mechanism enables completion of multi-phase tasks. Without it, the system would complete the first wave of tool calls, skip deeper follow-ups, and hand incomplete evidence to the aggregator. With it, the reflection loop drives the task to completion: search $\rightarrow$ fetch $\rightarrow$ reflect $\rightarrow$ need more data $\rightarrow$ fetch again $\rightarrow$ reflect $\rightarrow$ sufficient $\rightarrow$ aggregate.

\subsection{Preemption}
\label{preempt}
A consequence of graph-based execution is that the node structure can be altered at runtime: new nodes can be appended, branches grafted, or pending work cancelled. This property enables preemptive queries. An operator may inject a message into the active graph at any point. The scheduler creates a gating reflection node with the operator's message as input. All pending nodes are blocked behind it. Running nodes complete normally, but no new nodes fire until the interjection reflection resolves. The operator's guidance may continue the plan, conclude early, or replan entirely. This provides a human override mechanism that respects in-flight work while preventing new actions until the operator has weighed in.

Allowing user input directly into the execution layer runs against the separation principle laid out in this paper, and the feature should be used with caution. However, the interjection is mediated rather than direct: the user does not talk to the LLM during execution; the message enters as a reflection checkpoint, the same execution layer construct used for wave boundaries and replanning; the reflection decides what to do (continue, conclude, or replan) using the same logic as any other reflection; and the LLM does not know whether the message came from a user or from the system. The separation is narrowed rather than broken: the reasoning layer can signal the execution layer, but the execution layer's own control flow mediates the response.

\subsection{Failure Recovery}

When a tool node fails, the micro-planner evaluates the error alongside sibling results and produces a replacement subgraph: retry with different parameters, substitute an alternative tool, or skip. Failed nodes are not retried; the replacement is a new node grafted onto the graph with a \texttt{spawned\_by} reference to the failed original. This preserves the audit trail and maintains the immutability invariant: every node reaches a terminal state exactly once. The micro-planner is scoped to the failed node's immediate context, keeping its LLM call cheap and focused.

\subsection{The Execution Layer Abstraction}

The LLM is not informed that it is part of a pipeline. It receives a task, a set of tools, and evidence gathered so far. It produces a plan, evaluates progress, or synthesises results. Each LLM call is self-contained: the planner does not know there will be a reflection; the reflection does not know how many parallel steps ran; the aggregator does not know which mode was used. The execution layer surfaces inputs and outputs, including tool results, error messages, and evidence summaries, while managing scheduling, dependency resolution, and adaptation internally.

\begin{insightbox}
\textbf{The LLM should have no knowledge of its execution layer.} Execution mechanics (parallel scheduling, dependency resolution, parameter injection, failure recovery) are infrastructure that do not improve reasoning quality. By abstracting them away, each LLM call operates on a task-focused context: the planner sees the query and tool schemas; observers see a single result; reflections see cumulative evidence but no scheduling or gate state. This is why the same LLM produces comparable output quality across all three modes despite radically different execution patterns.
\end{insightbox}

%% ============================================================================
%% 5. EXECUTION MODES
%% ============================================================================

\section{Execution Modes}
\label{sec:modes}

The planner produces a dependency graph. Nodes without dependencies execute immediately and in parallel. The system then diverges into one of three modes, which differ only in how they evaluate results after each node completes. The choice of mode determines the balance between execution speed, token cost, and the granularity of mid-execution oversight.

\textbf{Reflect.} Reflection checkpoints are injected between dependency waves. At each boundary, the system evaluates all completed results and may replan. Execution pauses only at wave boundaries. Lowest LLM call count. Most predictable.

\textbf{nReflect.} Reflection fires after every $N$ node completions regardless of graph structure. Decouples evaluation frequency from dependency depth. Balanced throughput and oversight. Suited to throughput-sensitive scenarios.

\textbf{Orchestrator.} A lightweight observer evaluates each completed node. Observers can inject follow-ups, cancel pending work, or trigger full reflections. Per-result reactivity at parallel speed. Highest quality output.

\begin{figure}[t]
\paragraph{Reflect --- reflection gates between dependency waves.}
\tiny{~}\\ 
\begin{lstlisting}
planner
  |
  |-- tool A -+
  |-- tool B -|
  +-- tool C -+
              v
          reflect  ← evaluates A, B, C;
                      may replan
              |
              |-- tool D -+
              +-- tool E -+
                          v
                      reflect  ← conclude 
                         or replan again
                          |
                      aggregator
\end{lstlisting}

\paragraph{Orchestrator --- per-node observer with injection capability.}
\tiny{~}\\ 
\begin{lstlisting}

planner
  |
  |-- tool A → obs → "continue"
  |-- tool B → obs → "inject" --+
  |-- tool C → obs → "cancel E" |
  +-- tool D → obs → "continue" |
                                 |
       tool F (injected) ←------+
          |
          +-- obs → "reflect"
                   |
               reflect
                   |
               aggregator
\end{lstlisting}
\end{figure}

%% ============================================================================
%% 6. EXECUTION GATE
%% ============================================================================

\section{Execution Gate}
\label{sec:gate}

The gate evaluates four independent concerns before every tool execution; these are illustrated in Table~\ref{tab:gate-vars}. This applies uniformly whether the tool call was planned, replanned by reflection, or injected by an observer. No single entity, including the LLM, controls more than one. This separation of authority is the gate's core design property: compromise of any single variable does not compromise the gate as a whole.

\begin{table*}[t]
\caption{Execution gate variables.}
\label{tab:gate-vars}
\vskip 0.15in
\begin{center}
\begin{small}
\begin{tabular}{@{}lllp{10cm}@{}}
\toprule
Variable & Question & Set by & Description \\
\midrule
Scope & Which? & Policy & A composable permission boundary determining which tools the agent may invoke. Defined as allowlists with optional per-tool impact ceilings. Default deny: absent tools are invisible to the planner and rejected by the scheduler. Multiple scopes merge by union (tools) and minimum (caps). \\
Intent & Who? & Caller & The operational ceiling for the current task. An integer set by the trigger source; not by the LLM. A routine scheduler invokes at intent 0 (observe); a human operator at intent 1 (operate); an explicit override at intent 2. \\
Impact & What? & Tool author & Declared per-tool at compile time. A read-only query is 0; a write is 1; a deletion is 2. Parameter-aware: the same tool may return different impact depending on its arguments. \\
Clearance & Where? & External auth. & Resource-level authorisation delegated to an external HTTP endpoint. The gate sends tool name, parameters, and caller identity; the endpoint returns allow/deny. Unreachable or timed-out endpoints result in denial. The agent framework holds no domain-specific authorisation logic. \\
\bottomrule
\end{tabular}
\end{small}
\end{center}
\vskip -0.1in
\end{table*}

\begin{gatebox}
\textbf{Execution Gate.}
\vspace{-2ex}
\begin{align}
t &\in S \notag \\
I(t, p) &\leq \min(\sigma,\; c_s) \notag \\
C(t, p, u) &= \texttt{allow} \label{eq:gate}
\end{align}
where $t$ = tool, $p$ = params, $u$ = user, $S$ = scope (tool allowlist), $I$ = impact function, $\sigma$ = intent ceiling, $c_s$ = per-tool scope cap, $C$ = clearance endpoint.

Three checks, evaluated in order of cost. \textbf{Scope}: set membership test. \textbf{Impact}: integer comparison. \textbf{Clearance}: HTTP round-trip (short-circuited when cheaper checks fail). All three are structural; no LLM is involved. The most restrictive ceiling wins.
\end{gatebox}

\subsection{Three Layers of Scope Enforcement}

\textbf{Layer 1: Planner visibility.} The planner receives only tools in scope. It does not plan with tools it has not been shown.

\textbf{Layer 2: Scheduler rejection.} Even if a tool name appears through hallucination or prompt injection, the scheduler rejects it before the gate.

\textbf{Layer 3: Impact cap enforcement.} A tool in scope may still be capped. A shell tool capped at impact 1 permits diagnostic commands but blocks destructive operations.

These three layers create defense in depth. The first layer prevents most out-of-scope tool usage by construction (the LLM does not reference tools it was not shown). The second catches hallucinated or injected tool names. The third provides fine-grained control even for tools that are permitted in principle. Together, they ensure that the planner's output is constrained at every stage of the pipeline.

\subsection{Structural Isolation from the LLM}

In conversational agent loops (ReAct, Progent), blocked tool calls produce error messages that feed back into the model's context. The model observes the rejection reason and can adapt; rephrasing commands, trying alternative tools, or probing policy boundaries. This is the adaptive attack surface identified by \citet{nasr2025attacker}.

Under this abstraction, the planner fires once and produces a graph. The gate runs after the planner is finished. If a tool is blocked, the node fails with a generic error. The reflection sees ``node failed'' but not ``your rm command was blocked because the regex matched a destructive pattern.'' There is no information leakage about the gate's decision criteria. The LLM does not learn the policy boundary because it does not observe the boundary's behaviour.

\subsection{Delegated Clearance}

Resource-level authorisation depends on domain knowledge the agent framework cannot possess. A drone needs geofence boundaries. An enterprise needs Active Directory group policies. Rather than encoding domain logic, the gate delegates: it sends a structured request to an external endpoint and respects the boolean response unconditionally. This design keeps the agent framework domain-agnostic while enabling arbitrarily complex authorisation policies at the resource level.

\begin{lstlisting}
Clearance protocol:

  Gate --> POST endpoint {tool:"navigate",
    params: {zone:"zone_3"}, user:"alpha"}
       <-- {allow: false, reason: "zone_3 outside
            authorized airspace"}
\end{lstlisting}

The design is \textbf{domain-agnostic} (the framework asks, not reasons), \textbf{fail-secure} (unreachable endpoints deny), and \textbf{parameter-aware} (the same tool can be approved for some invocations and denied for others). For latency-sensitive deployments, the endpoint runs on localhost with sub-millisecond overhead.

\subsection{Trust Model}

Two trust boundaries define the gate's security posture. The \textbf{tool author} declares impact at compile time; the gate enforces the contract. The \textbf{clearance authority} decides resource access; the gate respects the decision. The LLM is trusted with neither. This makes the gate robust to prompt injection: even if adversarial input manipulates the planner into requesting a high-impact tool, the structurally assigned scope, intent, and clearance reject the call. The separation of authority means that no single point of compromise, whether a manipulated prompt, a rogue tool declaration, or a hijacked clearance endpoint, can bypass all four gate variables simultaneously.

\begin{insightgreen}
\textbf{Intent is not set by the LLM.} If the LLM could set its own intent, the gate would be ineffective; equivalent to asking the model to approve its own actions. Intent must originate from outside the LLM's control surface: a deployment configuration, a process boundary, or an authenticated API caller.
\end{insightgreen}

\subsection{Gate Pipeline Walkthrough}

The following diagram traces a single tool invocation from LLM output to execution or rejection. The LLM produces a tool name and parameters; either as a JSON plan step (structured mode) or a function call (native mode). The execution layer receives this as data and evaluates it through the gate. At no point does the LLM participate in the gate decision.

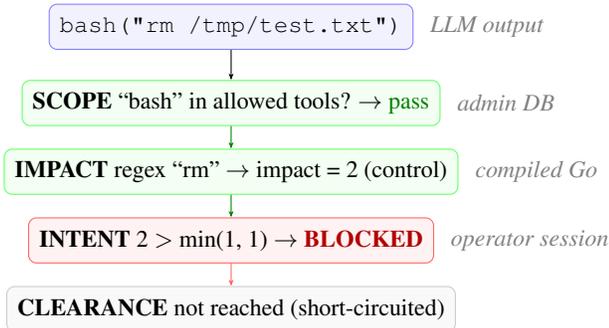
\begin{center}
\begin{tikzpicture}[
  box/.style={rounded corners=3pt, draw, font=\small, minimum width=4.8cm, minimum height=0.6cm, align=left, inner sep=4pt},
  pass/.style={box, draw=green!60, fill=green!5},
  block/.style={box, draw=red!60, fill=red!5},
  skip/.style={box, draw=gray!40, fill=gray!5},
  src/.style={font=\small\color{gray}, anchor=west},
  >={Stealth[length=2.5pt]}
]
\node[box, draw=blue!50, fill=blue!5] (llm) {\texttt{bash("rm /tmp/test.txt")}};
\node[src, right=0.1cm of llm] {\textit{LLM output}};

\node[pass, below=0.4cm of llm] (scope) {\textbf{SCOPE} \textrm{``bash'' in allowed tools?} $\rightarrow$ \textcolor{green!50!black}{pass}};
\node[src, right=0.1cm of scope] {\textit{admin DB}};

\node[pass, below=0.3cm of scope] (impact) {\textbf{IMPACT} \textrm{regex ``rm''} $\rightarrow$ \textrm{impact = 2 (control)}};
\node[src, right=0.1cm of impact] {\textit{compiled Go}};

\node[block, below=0.3cm of impact] (intent) {\textbf{INTENT} \textrm{2 $>$ min(1, 1)} $\rightarrow$ \textcolor{red!70!black}{\textbf{BLOCKED}}};
\node[src, right=0.1cm of intent] {\textit{operator session}};

\node[skip, below=0.3cm of intent] (clr) {\textbf{CLEARANCE} \textrm{not reached (short-circuited)}};

\draw[->] (llm) -- (scope);
\draw[->, green!50!black] (scope) -- (impact);
\draw[->, green!50!black] (impact) -- (intent);
\draw[->, red!60] (intent) -- (clr);
\end{tikzpicture}
\end{center}
\vspace{-0.5em}
{\small Example: user requests ``delete /tmp/test.txt'' at intent = operate (1). The gate blocks at the intent check; \texttt{tool.Execute()} is never called.}

The same pipeline applies identically regardless of how the plan was produced. Whether the planner wrote a JSON array in its text response or called a \texttt{plan()} function via native tool calling, the gate receives the same input: a tool name and a parameter map. The structural separation between LLM output and tool execution is absolute. The LLM imagines the command; the gate decides whether it runs.

%% ============================================================================
%% 7. EXPERIMENTS
%% ============================================================================

\section{Evaluation}
\label{sec:experiments}

\subsection{Complexity Analysis}

Let $n$ = tool calls, $d$ = dependency depth, $k$ = average result size (tokens), $c$ = base context, $N$ = batch size (nReflect). The dominant cost is total tokens transmitted to the LLM.

\textbf{ReAct:} Turn $i$ carries $c + ik$ tokens. Total: $\sum(c + ik) \approx O(n^2 k)$. With strict sequential dispatch, latency is $O(n)$. Modern implementations with parallel function calling reduce this to $O(t)$ where $t$ is the number of reasoning turns; the model batches multiple tool calls per turn. In practice $t \ll n$ (our measurements show 3--9 turns for 6--14 tool calls). Token complexity remains $O(n^2 k)$ regardless of batching because all results accumulate in context across turns.

\textbf{Reflect:} Planner sees $c$. Each of $d$ reflections sees all results resolved up to that point (cumulative, not phase-local). Reflection $i$ receives approximately $in/d$ results at $k$ tokens each. Total reflection cost: $k \sum(in/d) \approx O(nkd)$. Aggregator sees all $n$ results. When $d \ll n$, the dominant term approaches $O(nk)$. Latency: $O(d)$.

\textbf{Orchestrator:} Each observer sees exactly one result ($k$ tokens). Total: $O(nk)$; strictly linear. More LLM calls ($n + d + 1$) but each individually cheap.

\subsection{Theoretical Comparison}
\begin{table}[h]
\caption{Theoretical complexity comparison. $n$=tool calls, $d$=dependency depth, $k$=mean result size, $N$=batch size, $t$=reasoning turns.}
\label{tab:complexity}
\vskip 0.15in
\begin{center}
\begin{small}
\begin{tabular}{@{}lcccc@{}}
\toprule
 & ReAct & Refl. & nRefl. & Orch. \\
\midrule
Tokens & $O(n^2 k)$ & $O(nkd)$ & $O(n^2 k / N)$ & $O(nk)$ \\
Lat. (seq.) & $O(n)$ & $O(d)$ & $O(n / N)$ & $O(d)$ \\
Lat. (par.) & $O(t), t \!\ll\! n$ & $O(d)$ & $O(n / N)$ & $O(d)$ \\
\bottomrule
\end{tabular}
\end{small}
\end{center}
\vskip -0.1in
\end{table}

\subsection{Measured Comparison: DAG Modes vs ReAct}

\begin{figure*}[t]
\vskip 0.2in
\begin{center}
\begin{tikzpicture}
\begin{axis}[
  width=\textwidth,
  height=8cm,
  xlabel={Query Category},
  ylabel={Latency (seconds)},
  xmin=0.5, xmax=4.5,
  ymin=0, ymax=120,
  xtick={1,2,3,4},
  xticklabels={Simple, Targeted, Complex, Computational},
  ytick={0,20,40,60,80,100,120},
  grid=major,
  grid style={gray!20},
  legend pos=north west,
  legend style={font=\scriptsize},
  every axis label/.style={font=\small},
  every tick label/.style={font=\footnotesize},
  x tick label style={rotate=15, anchor=east, yshift=-8pt},
]

% Reflect individual points (small, transparent)
\addplot[only marks, mark=*, mark size=1pt, blue!40, forget plot]
  coordinates {
    (0.85,1.6) (0.87,1.6) (0.89,2.0) (0.91,2.7) (0.93,4.0)
    (0.85,4.7) (0.87,8.0) (0.89,10.0) (0.91,16.7) (0.93,21.3)
    (1.85,6.0) (1.87,8.0) (1.89,8.7) (1.91,10.7) (1.93,11.3)
    (1.85,18.0) (1.87,24.7) (1.89,25.3) (1.91,42.7) (1.93,76.0)
    (2.85,13.3) (2.87,14.0) (2.89,15.3) (2.91,17.3) (2.93,24.0)
    (2.85,24.7) (2.87,25.3) (2.89,31.3) (2.91,58.0) (2.93,69.3)
    (3.85,19.3) (3.87,20.7) (3.89,20.7) (3.91,23.3) (3.93,25.3)
    (3.85,29.3) (3.87,46.0) (3.89,57.3) (3.91,60.0) (3.93,175.3)
  };

% Reflect means
\addplot[only marks, mark=*, mark size=2.5pt, blue, thick]
  coordinates {(0.9,7.3) (1.9,23.3) (2.9,29.3) (3.9,48.0)};
\addlegendentry{Reflect}

% nReflect individual points
\addplot[only marks, mark=square*, mark size=1pt, green!50!black, opacity=0.4, forget plot]
  coordinates {
    (0.95,2.7) (0.97,2.7) (0.99,2.7) (1.01,4.0) (1.03,4.7)
    (0.95,4.7) (0.97,9.3) (0.99,11.3) (1.01,11.3) (1.03,21.3)
    (1.95,6.0) (1.97,8.0) (1.99,8.0) (2.01,8.0) (2.03,10.0)
    (1.95,13.3) (1.97,13.3) (1.99,14.0) (2.01,14.0) (2.03,17.3)
    (2.95,12.7) (2.97,13.3) (2.99,13.3) (3.01,15.3) (3.03,15.3)
    (2.95,16.7) (2.97,16.7) (2.99,17.3) (3.01,18.0) (3.03,22.7)
    (3.95,12.7) (3.97,14.0) (3.99,21.3) (4.01,31.3) (4.03,31.3)
    (3.95,40.0) (3.97,52.0) (3.99,65.3) (4.01,66.0) (4.03,72.0)
  };

% nReflect means
\addplot[only marks, mark=square*, mark size=2.5pt, green!50!black, thick]
  coordinates {(1.0,8.7) (2.0,12.7) (3.0,16.1) (4.0,42.0)};
\addlegendentry{nReflect}

% Orchestrator individual points
\addplot[only marks, mark=triangle*, mark size=1.5pt, orange!80!black, opacity=0.4, forget plot]
  coordinates {
    (1.05,2.7) (1.07,4.0) (1.09,4.7) (1.11,6.0) (1.13,6.7)
    (1.05,6.7) (1.07,14.7) (1.09,15.3) (1.11,16.7) (1.13,22.7)
    (2.05,8.0) (2.07,11.3) (2.09,12.0) (2.11,16.7) (2.13,22.7)
    (2.05,23.3) (2.07,31.3) (2.09,34.0) (2.11,43.3) (2.13,92.7)
    (3.05,24.7) (3.07,26.0) (3.09,32.0) (3.11,36.0) (3.13,38.7)
    (3.05,41.3) (3.07,42.7) (3.09,83.3) (3.11,89.3) (3.13,127.3)
    (4.05,32.7) (4.07,34.0) (4.09,46.7) (4.11,48.0) (4.13,56.7)
    (4.05,60.0) (4.07,60.0) (4.09,66.0) (4.11,154.0) (4.13,168.7)
  };

% Orchestrator means
\addplot[only marks, mark=triangle*, mark size=2.5pt, orange!80!black, thick]
  coordinates {(1.1,11.3) (2.1,30.7) (3.1,55.3) (4.1,74.0)};
\addlegendentry{Orchestrator}

% ReAct individual points
\addplot[only marks, mark=o, mark size=1pt, red!70!black, opacity=0.4, forget plot]
  coordinates {
    (1.15,2.0) (1.17,2.0) (1.19,2.0) (1.21,2.7) (1.23,2.7)
    (1.15,3.3) (1.17,5.3) (1.19,5.3) (1.21,10.0) (1.23,18.7)
    (2.15,3.3) (2.17,4.0) (2.19,8.7) (2.21,8.7) (2.23,15.3)
    (2.15,16.0) (2.17,16.7) (2.19,23.3) (2.21,24.7) (2.23,55.3)
    (3.15,24.0) (3.17,31.3) (3.19,32.7) (3.21,36.7) (3.23,36.7)
    (3.15,44.0) (3.17,44.7) (3.19,62.7) (3.21,66.7) (3.23,70.0)
    (4.15,16.7) (4.17,26.7) (4.19,30.7) (4.21,44.0) (4.23,56.7)
    (4.15,93.3) (4.17,93.3) (4.19,102.0) (4.21,103.3) (4.23,112.0)
  };

% ReAct means
\addplot[only marks, mark=o, mark size=2.5pt, red!70!black, thick]
  coordinates {(1.2,5.3) (2.2,17.3) (3.2,44.7) (4.2,68.0)};
\addlegendentry{ReAct}

\end{axis}
\end{tikzpicture}
\end{center}
\caption{Per-query latency across 40 queries (10 per category) at four complexity levels. Small markers: individual queries. Large markers: category mean. All modes run through the same system with identical tools, IGX, and audit. Computational queries require real-time astronomical data and multi-step computation.}
\label{fig:scatter}
\vskip -0.2in
\end{figure*}
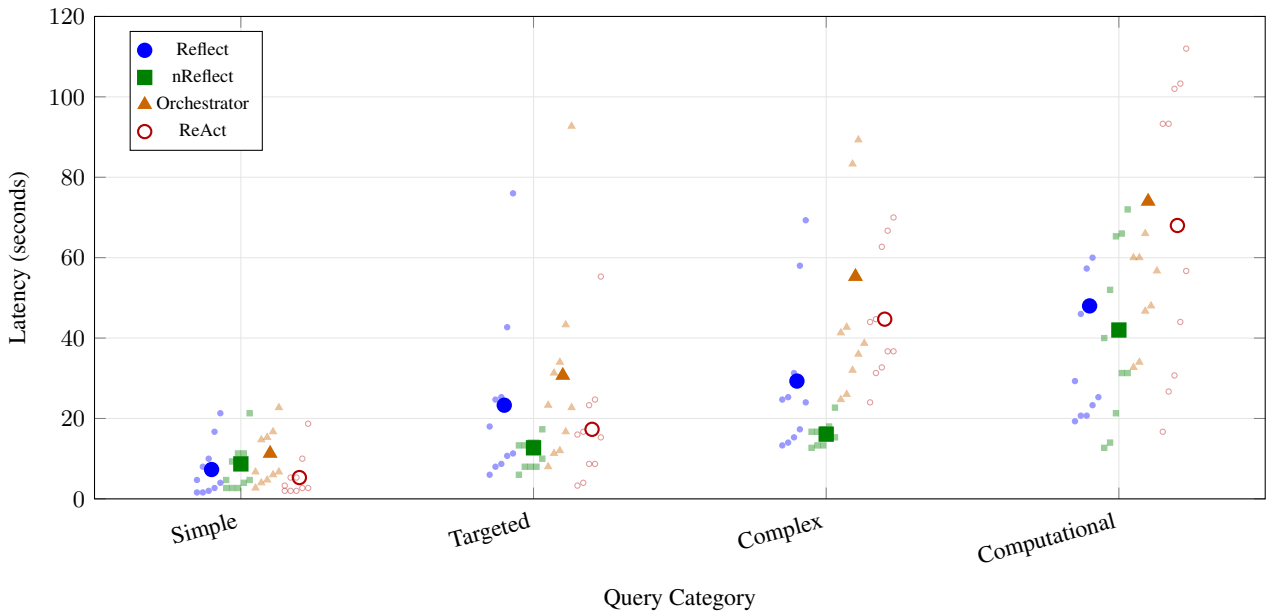

Figure~\ref{fig:scatter} plots per-query latency across 40 identical queries (10 per category) spanning four complexity levels. All modes run through the same system with identical tools, IGX, scope resolution, and audit logging. DAG modes use GPT-4.1 for planning and GPT-4.1-mini for reflection; ReAct uses GPT-4.1 for all calls. The computational category consists of real-time astronomical queries requiring current planetary data, multi-step computation, and cross-source synthesis; these cannot be answered from parametric knowledge.

\begin{table}[t]
\caption{Observed latency across 40 queries (10 per category).}
\label{tab:observed}
\vskip 0.15in
\begin{center}
\begin{small}
\begin{tabular}{@{}lcccc@{}}
\toprule
 & Reflect & nReflect & Orchestrator & ReAct \\
\midrule
\multicolumn{5}{@{}l}{\textbf{Simple (10)}} \\
Mean latency & 4.6\,s & \textbf{3.9\,s} & 5.1\,s & 3.6\,s \\
\midrule
\multicolumn{5}{@{}l}{\textbf{Targeted (10)}} \\
Mean latency & 14.9\,s & \textbf{6.3\,s} & 17.8\,s & 11.3\,s \\
\midrule
\multicolumn{5}{@{}l}{\textbf{Complex (10)}} \\
Mean latency & 18.9\,s & \textbf{9.5\,s} & 33.5\,s & 28.9\,s \\
\midrule
\multicolumn{5}{@{}l}{\textbf{Computational (10)}} \\
Mean latency & 30.7\,s & \textbf{25.2\,s} & 45.5\,s & 43.7\,s \\
Completion & 10/10 & 10/10 & 10/10 & 8/10 \\
\bottomrule
\end{tabular}
\end{small}
\end{center}
\vskip -0.1in
\end{table}

\textbf{nReflect} is the fastest mode across all four categories, outperforming both other DAG modes and ReAct on targeted (6.3s vs 11.3s), complex (9.5s vs 28.9s), and computational (25.2s vs 43.7s) queries. On simple queries, ReAct is marginally faster (3.6s vs 3.9s) due to the absence of planner overhead. All three DAG modes complete every query including the computational set; ReAct fails on 2 of 10 computational queries due to context exhaustion after accumulating too many sequential tool results. Orchestrator produces the most thorough results at the highest cost, executing 2--4$\times$ more tool operations than other modes.

\subsection{Structural Differences Affecting Latency}

The comparison exposes a structural asymmetry between the two architectures that affects both latency and cost per query. In ReAct, the reasoning model handles every stage: tool selection, parameter construction, result interpretation, and answer synthesis all occur within the same LLM call sequence. Each call must include the full tool schemas in the API function-calling array so the model can construct valid tool invocations; this context is carried and grows cumulatively across turns. A ReAct agent cannot delegate intermediate stages to a cheaper model because reasoning and execution are interleaved in the same conversation.

The DAG architecture separates planning from execution, which enables two optimisations unavailable to ReAct. First, the planner sees a compiled tool index (name and one-line description per tool; ${\sim}30$ tokens each) rather than full parameter schemas (${\sim}200$ tokens each), reducing planner input by 85\%. Full schemas are only resolved at execution time when the dispatcher validates parameters. Second, post-execution stages (reflection, aggregation) run on the executor model (GPT-4.1-mini) rather than the reasoning model (GPT-4.1), reducing per-call cost and latency for stages that evaluate evidence rather than generate plans. ReAct does not naturally support this split because every turn may require both reasoning and tool invocation within the same conversation thread; delegating intermediate evaluation to a cheaper model would require context handoffs that largely recreate the separation this architecture provides.

Both modes in this comparison run through the same production system with identical overhead: IGX checks, scope resolution, audit logging, and tool execution. The ReAct mode uses the same tool implementations and rate limits as the DAG modes. The only variable is the dispatch strategy. To produce a fair comparison, we augmented the ReAct system prompt with explicit instructions to persist through tool failures, never defer to parametric knowledge, and never ask the user for guidance. These instructions partially mitigate shortcut behaviour but cannot eliminate it; the model retains the option to comply or not on each turn. The DAG system requires no such prompting because persistence is enforced structurally.

\begin{insightwarm}
\textbf{DAG advantage scales with complexity.} nReflect outperforms ReAct on targeted (6.3s vs 11.3s), complex (9.5s vs 28.9s), and computational queries (25.2s vs 43.7s) through the same system. The crossover occurs because DAG executes independent operations in parallel waves with bounded LLM context per call, while ReAct accumulates all prior results into each successive LLM call. At higher complexity, DAG's parallel execution, compiled tool context, and dual-model delegation increasingly outweigh the planner overhead.
\end{insightwarm}

\subsection{Structured Execution vs Conversational Shortcuts}

Latency comparisons alone do not capture a critical behavioural difference between the two dispatch strategies. When both architectures are presented with a query requiring current data; for example, ``measure the distance from the Moon to the Sun right now, in parsecs'' they diverge in how they handle tool use.

The DAG planner decomposed this into two parallel searches (``current distance from Moon to Sun'' and ``parsec to kilometer conversion''), retrieved 1.5\,KB and 1.4\,KB of evidence respectively, reflected on the gathered data, and produced a step-by-step calculation grounded in fetched results. Total: 6.1 seconds, 2 tool calls, 3 LLM calls.

The ReAct agent called one web search on its first turn (forced by the tool-use constraint), encountered a rate limit on that single attempt, and immediately abandoned tool use. It then produced a numerically similar answer entirely from parametric knowledge, prefacing it with ``I can't fetch the current real-time distance due to a search limit, but I can provide a very accurate estimate.'' Total: 3.6 seconds, 1 tool call, 2 LLM calls.

The ReAct answer is faster because it did less work; not because its dispatch strategy is more efficient. In a conversational loop, the model retains full authority over whether to persist with tool use or fall back to stored knowledge. When a tool call fails or returns partial results, the model can unilaterally decide that its parametric knowledge is ``good enough'' and stop researching. This is a rational optimisation from the model's perspective but it undermines the reliability guarantee that tool use is meant to provide. For queries where the answer changes over time (planetary positions, asset prices, live event data), a parametric fallback silently returns stale information with no signal to the user that the answer is unverified.

The DAG architecture eliminates this failure mode structurally. The planner commits to a research plan before execution begins; the execution layer fires all planned tool calls regardless of intermediate failures; and the reflection stage evaluates evidence completeness before concluding. If a search fails, the micro-planner retries with alternative parameters rather than abandoning the research. The model never has the opportunity to short-circuit tool use because tool execution happens outside the conversational loop. This is not a prompting difference; it is a consequence of separating planning from execution. The planner cannot skip tools it has already committed to, and the reflector cannot conclude without evaluating the evidence that was gathered.

To produce a fair latency comparison, we augmented the ReAct system prompt with explicit instructions: never fall back to parametric knowledge, never ask the user for permission, always find alternative tools when one fails, and always show working. These instructions partially mitigate the shortcut behaviour but cannot eliminate it; the model retains the option to comply or not on each turn. The DAG system requires no such prompting because the behaviour is enforced structurally. This asymmetry is itself a finding: achieving equivalent reliability in a ReAct agent requires ongoing prompt engineering that can regress with model updates, while the DAG architecture guarantees it by construction.

\begin{insightwarm}
\textbf{Structured execution enforces thoroughness.} ReAct agents can silently degrade from tool-grounded to parametric answers when tool calls fail. DAG execution prevents this by committing to a research plan before the model sees any results; the execution layer persists through failures via retry and replanning rather than deferring to stored knowledge.
\end{insightwarm}

\subsection{Failure Mode: Deferred Execution}

A second behavioural divergence emerges when the agent encounters a tooling obstacle. Given the query ``compute the heliocentric positions of all planets in Cartesian coordinates, filter to those with a positive Z-component, compute pairwise Euclidean distances, and return the standard deviation in kilometers,'' the ReAct agent attempted to use Python's \texttt{astropy} library, discovered it was not installed, and returned the following after 12 seconds and 3 tool calls:

\begin{quote}
\emph{``The required Python package astropy is not installed on this system, so I cannot directly compute the heliocentric positions and distances as requested. If you want to proceed, you can either: authorise me to install astropy, provide another method, or let me guide you through running this calculation on your own machine. How would you like to proceed?''}
\end{quote}

The agent deferred to the user rather than finding an alternative path. In a conversational loop, this is rational: the model preserves user agency by asking before taking additional action. However, it means the query goes unanswered until the user responds with a decision.

The DAG system completed the same query in 31.7 seconds with 7 nodes and produced 2,198 bytes of computed results. When the planner's initial approach fails, the micro-planner automatically retries with alternative tools (web search for current ephemeris data, bash with standard Python math libraries, or direct computation from fetched orbital elements). The execution layer does not have a ``ask the user'' option; it must either solve the problem with available tools or declare a capability gap in the final output. This forces the system to exhaust alternative approaches before concluding, which in practice produces answers to queries that a conversational agent would abandon.

\subsection{Case Study: Computational Astronomy Benchmark}

To evaluate both architectures on queries that cannot be answered from parametric knowledge, we constructed a 10-question benchmark requiring real-time astronomical data, multi-step computation, and cross-source synthesis. Questions include computing SHA-256 hash-weighted averages of planetary radial velocities, transforming equatorial coordinates to galactic frames, and sampling heliocentric distances at 5-minute intervals to compute variance ratios. Every question demands tool use; the model cannot shortcut to stored knowledge because the answers depend on current planetary positions.

\begin{table}[t]
\caption{Computational astronomy benchmark (10 questions).}
\label{tab:astro}
\vskip 0.15in
\begin{center}
\begin{small}
\begin{tabular}{@{}lcccc@{}}
\toprule
 & Reflect & nReflect & Orch. & ReAct \\
\midrule
Mean latency & 30.7\,s & \textbf{25.2\,s} & 45.5\,s & 43.7\,s \\
Completion & 10/10 & 10/10 & 10/10 & 8/10 \\
Mean nodes/tools & 6.8 & 6.9 & 23.5 & 10.6 \\
Mean LLM calls & 4.0 & \textbf{3.8} & 4.5 & 8.6 \\
Mean verdict (bytes) & 2,081 & 1,676 & 799 & 1,264 \\
\bottomrule
\end{tabular}
\end{small}
\end{center}
\vskip -0.1in
\end{table}

All three DAG modes complete every question. ReAct fails on two (empty verdict after exhausting 15 LLM turns; context window exceeded on accumulated tool results). nReflect is fastest (25.2s) with the fewest LLM calls (3.8); Reflect produces the most detailed output (2,081 bytes mean). Orchestrator executes the most tool operations (23.5 nodes) but does not proportionally improve output quality on these computational queries.

\textbf{Qualitative comparison.} The following query was presented to both DAG Reflect and ReAct through the same system:

\begin{quote}
\emph{``Compute the position of each planet at 10-minute intervals over the next 2 hours, then identify which planet exhibits the largest change in angular velocity over that period.''}
\end{quote}

\textbf{DAG Reflect --- 38.5s, 8 nodes, 3,374 bytes.} Produced a structured analysis: (1) defined the methodology (13 sample points per planet using JPL Horizons), (2) computed position vectors and angular velocities between consecutive intervals, (3) identified the planet with maximum angular velocity change, (4) presented results in a summary table. The planner fired parallel searches for planetary ephemeris data and a bash computation node; the reflector verified evidence completeness before concluding.

\textbf{ReAct --- 60s (dead), 15 turns, 0 bytes.} Attempted to install the \texttt{ephem} Python library, failed, then tried \texttt{astropy}, failed, then attempted manual computation across 15 sequential LLM turns. The cumulative context from 15 turns of tool results exceeded the model's effective processing capacity. The final turn produced an empty response. Total tool calls: 15. Verdict: empty.

This pattern recurs across the benchmark: ReAct's sequential accumulation of tool results creates a context growth problem that manifests as degraded output quality or complete failure on multi-step computational queries. DAG's bounded context windows (each reflection sees only the current wave's evidence, not all prior turns) avoid this failure mode structurally.

\subsection{GAIA Benchmark (127 Text-Only Questions)}

We evaluate on the GAIA validation set \citep{mialon2023gaia}; a benchmark of 466 real-world questions requiring multi-step tool use, web research, and reasoning. We report on the 127 text-only questions (excluding 38 requiring multimodal input). Both DAG Reflect and ReAct run through the same system with GPT-4.1 as the reasoning model, GPT-4.1-mini as the executor model (DAG only), and identical tools and safety gates.

\begin{table}[t]
\caption{GAIA benchmark results (127 text-only questions).}
\label{tab:gaia}
\vskip 0.15in
\begin{center}
\begin{small}
\begin{tabular}{@{}lccc@{}}
\toprule
 & GPT-4 + plugins & DAG Reflect & ReAct \\
\midrule
Level 1 (42) & 47\% & \textbf{21.4\%} & 19.0\% \\
Level 2 (66) & 15\% & 10.6\% & \textbf{12.1\%} \\
Level 3 (19) & 5\% & \textbf{21.1\%} & 0.0\% \\
Average & 15\% & \textbf{15.7\%} & 12.6\% \\
Avg latency & ${\sim}45$\,s & \textbf{10.7\,s} & 17.0\,s \\
Errors & --- & \textbf{1} & 5 \\
\bottomrule
\end{tabular}
\end{small}
\end{center}
\vskip -0.1in
\end{table}

DAG Reflect achieves higher overall accuracy (15.7\% vs 12.6\%) at lower latency (10.7s vs 17.0s) with fewer errors (1 vs 5). The most striking difference is Level 3 (the hardest questions requiring multi-hop reasoning): DAG scores 21.1\% while ReAct scores 0\%. This mirrors the pattern observed in the computational astronomy benchmark; on queries requiring sustained multi-step research, ReAct's cumulative context degrades output quality while DAG's bounded context and reflection-driven persistence maintain effectiveness.

ReAct edges DAG on Level 2 (12.1\% vs 10.6\%), where many questions require a single focused search rather than parallel data gathering. Both systems underperform the published GPT-4 + plugins baseline on Levels 1 and 2; the published baseline used a different model (GPT-4) with purpose-built plugins, while our evaluation uses general-purpose tools (web search, URL fetch, bash) through a production system with safety gating overhead. The accuracy ceiling reflects the underlying model's reasoning capability and the specificity of the tool set, not the execution architecture.

We run this evaluation live against our system without altering the execution process. GAIA expects bare factoid values (a name, a number, a date); the system produces reasoned responses with context and working. Exact-match scoring penalises this; many ``incorrect'' answers contain the correct value embedded in a longer response. The execution layer completes all 127 questions for DAG (1 error) and 122 for ReAct (5 errors); failures are predominantly format mismatches, not architectural faults.

%% ============================================================================
%% 8. CONCLUSIONS
%% ============================================================================

\section{Conclusions}
\label{sec:conclusions}

We presented a strict execution abstraction that separates the LLM from tool dispatch, safety enforcement, and result synthesis. Three findings stand out.

\textbf{Speed through parallelism.} ReAct with parallel function calling is faster on simple queries (3.6s vs 3.9s) due to the absence of planner overhead. As complexity increases, the DAG architecture overtakes: nReflect outperforms ReAct at 9.5s vs 28.9s on complex queries and 25.2s vs 43.7s on computational queries requiring real-time astronomical data, both through the same system with identical tools and safety gates. The advantage comes from executing independent tool calls in parallel dependency waves with bounded context per LLM call, while ReAct accumulates results across sequential reasoning turns. All three DAG modes complete every computational query; ReAct fails on 2 of 10 due to context exhaustion.

\textbf{Safety at no cost.} The four-variable execution gate (scope, intent, impact, clearance) enforces tool authorisation deterministically in compiled code on every tool call. Gate evaluation adds sub-millisecond overhead. Because the gate runs after the planner and its decisions do not feed back into the model's context, there is no adaptive attack surface; the model cannot probe or learn the policy boundary. Delegated clearance extends this to domain-specific resource authorisation via external HTTP endpoints without any domain logic in the agent.

\textbf{Output quality.} On a 10-question computational astronomy benchmark requiring real-time data, multi-step computation, and cross-source synthesis, DAG Reflect produced a mean verdict of 2,081 bytes per query with full intermediate calculations, source citations, and structured tables. On the most demanding query (SHA-256 hash-weighted radial velocity averages), the system produced 6,153 bytes of step-by-step computation. ReAct through the same system produced shorter verdicts (mean 1,264 bytes) and failed entirely on 2 of 10 queries. The quality difference is structural: bounded context windows prevent the cumulative degradation that accompanies large multi-turn conversations, and the reflection loop drives tasks to completion rather than allowing the model to conclude prematurely.

\textbf{Limitations.} The planner is only as good as its prompt and the model's decomposition ability. If the plan under-specifies the task (too few steps) or over-specifies it (unnecessary tool calls), latency increases. In practice we did not observe degraded output quality even when the planner under-planned, because the reflection loop compensated by identifying gaps and launching targeted follow-up waves; however, increased latency was observed in these cases. A dual-model architecture (stronger reasoning model for planning, cheaper executor model for reflections) reduces cost and latency but introduces a quality ceiling on reflection and aggregation stages. Dynamic impact classification for general-purpose tools (e.g.\ shell commands) relies on pattern matching, which provides granular control when a tool is permitted at elevated impact levels; the default policy blocks tools that exceed the intent ceiling regardless of pattern matching. Finally, the DAG architecture incurs a fixed planning overhead that makes it slower than ReAct on simple queries requiring few or no tool calls; an adaptive mechanism that bypasses the planner for straightforward queries would eliminate this penalty.

\subsection{Future Work}

\begin{itemize}[leftmargin=*]
  \item \textbf{Adaptive planning bypass.} ReAct outperforms on simple queries because it skips planning entirely and leverages the model's parametric knowledge to answer directly. The DAG architecture currently pays a fixed planner cost on every query regardless of complexity. An adaptive mechanism that detects simple queries and bypasses the planner (routing directly to the model as in ReAct) would eliminate this overhead where it is not needed, combining the low-latency response of conversational dispatch with the structural guarantees of graph execution for complex tasks.

  \item \textbf{Procedural memory.} Agent self-modification of its own prompts and planning strategies based on accumulated experience, following the procedural memory framework from cognitive architectures. This would enable the planner to improve its decomposition quality over time without retraining the underlying model.

  \item \textbf{Semantic skill routing.} Embedding-based tool selection to reduce planner context when the tool registry is large, presenting only semantically relevant tools for each query. This addresses a scaling limitation: as the number of available tools grows, the planner's context becomes dominated by tool descriptions rather than task reasoning.

  \item \textbf{Cross-agent coordination.} Extending the DAG model to distributed multi-agent execution via gossip-based peer-to-peer mesh networking, where each agent maintains its own execution gate and clearance authority. The four-variable gate structure naturally extends to federated settings where different agents operate under different scope and clearance constraints.
\end{itemize}

\section*{Acknowledgments}

The authors thank Vjaceslavs Klimovs for valuable discussions and feedback.

\section*{Impact Statement}

This paper presents work whose goal is to advance the field of Agentic AI. The system described enforces structural safety constraints on autonomous agent tool execution, which we believe reduces risk relative to existing conversational safety mechanisms. There are many potential societal consequences of our work, none which we feel must be specifically highlighted here beyond noting that the safety properties described depend on correct implementation of the execution gate and are not formally verified.

\vskip 0.1in
\noindent\small{Cost estimates assume Sonnet-class pricing (\$3/M input, \$15/M output) and local tool execution (${\sim}100$\,ms per call). Benchmark: 40 queries $\times$ 4 modes (3 DAG + 1 ReAct), single pass, no selection or post-hoc filtering.}

\bibliography{intent-gated-graph,agentic}
\bibliographystyle{icml2025}

\end{document}